\documentclass[a4paper, amsfonts, amssymb, amsmath, reprint, showkeys, nofootinbib, twoside]{revtex4-1}
\usepackage[english]{babel}
\usepackage[utf8]{inputenc}
\usepackage[colorinlistoftodos, color=green!40, prependcaption]{todonotes}
\usepackage{amsthm}
\usepackage{mathtools}
\usepackage{physics}
\usepackage{xcolor}
\usepackage{graphicx}
\usepackage[left=23mm,right=13mm,top=35mm,columnsep=15pt]{geometry} 
\usepackage{adjustbox}
\usepackage{placeins}
\usepackage[T1]{fontenc}
\usepackage{lipsum}
\usepackage{csquotes}
\usepackage[pdftex, pdftitle={Article}, pdfauthor={Author}]{hyperref} 
\begin{document}
\title{The steady state distribution for diffusion \\ in a logarithmic-harmonic potential with stochastic resetting}

\hypersetup{hidelinks}

\author{Özgür Gültekin}
    \email{gultekino@yahoo.com}
       \affiliation{Kadikoy Moda 34710, Istanbul Turkey}

\date{\today} 

\begin{abstract}

The steady state distribution of the position of a Brownian particle diffusing in logarithmic-harmonic potential with stochastic resetting is obtained analytically. We show that there are two critical conditions that determine the behavior of the stationary distribution function (SDF). We also investigate how the steady state distribution, which occurs due to the nature of the logarithmic-harmonic potential in the absence of reset, changes in the presence of the reset mechanism. 

\end{abstract}

\maketitle

\section{INTRODUCTION} \label{sec:section1}

Stochastic reset processes, based on stopping a system at random intervals and restarting it in a given state, are commonly found in our environment \cite{Evans2020}. For example, animals looking for food in nature return to their nests after a while and then leave their nests again to continue the search \cite{Nathan2008}. Similar processes are observed in people's individual or social mobility \cite{Chen2022}. These can be seen as natural stochastic reset processes. It is known that stochastic reset or restart processes provide some advantages in finding the target, especially in the context of search algorithms \cite{Montanari2002,Bartumeus2009,Tong2008}. Stochastic reset can reduce the average time required to find a predetermined target \cite{Kusmierz2015}. In this case there is generally a reset rate $r$ that minimizes the mean time \cite{Evans2020}. That is, the stochastic reset mechanism can make search processes more efficient and can be used to optimize search algorithms \cite{Reuveni2016,Chechkin2018,Pal2019}.

Stochastic reset is a process that is effective in many different scientific fields such as intracellular diffusion in biology \cite{Bresslo}, in protein-DNA interactions \cite{Ghosh2018,Mirny2009,Shvets2016}, in the effect of antiviral treatments on the development of drug resistance \cite{Ramoso2020}, in ecology \cite{Evans2022}, in disaster (earthquakes, epidemics) modeling \cite{Visco2010,Plata2020}, in analysing of the sudden collapse of a financial entity \cite{Stojkoski2022,Montero2022}, in chemistry Michaelis Menten style reaction schemes \cite{Reuveni2014,Ray2021,Rotbart2015}, in some quantum systems \cite{Mukherjee2018,Sevilla2023}.

Evans and Majumdar, in their article published in 2011 \cite{Evans2011}, discussed a simple diffusion process in which a Brownian particle is sent to its initial position at a random moment by a reset mechanism. This article is known as the first article in the literature to draw attention to the effects of stochastic reset on the diffusion process. In the following years, the stochastic reset mechanism became an important research topic in non-equilibrium statistical mechanics due to its comprehensive and interesting theoretical features and its rich applications, despite its simple structure.

It is known that in a situation where a particle in Brownian motion is reset to its initial position at a constant rate $r$, after a sufficiently long time the system can reach a non-equilibrium steady state \cite{Evans2011,Ahmad2019}. Resetting throw the system out of balance and can still cause a steady state to occur \cite{Singh2022}. Therefore, the stochastic reset mechanism is a natural process to create an unbalanced steady state. A detailed investigation of such systems under a limiting potential has been made and it has been shown that the system exhibits a phase transition while relaxing to steady state \cite{Singh2020}.

The steady state of a Brownian particle diffusing at a random potential under the stochastic reset mechanism was obtained analytically for the mod potential and the harmonic potential, and it was revealed that the steady states differ depending on the nature of the potential \cite{Pal2015}. Analytical solutions can only be obtained for some simple potentials. Another important effect of stochastic resetting has to do with the first passage properties of a diffusing particle. Resetting can speed up or delay the first passage process \cite{DeBruyne2020}. By considering the effects of resetting in logarithmic potential on diffusion, it has been shown that the system exhibits different behavior depending on the ratio between the power of the potential and the thermal energy \cite{Ray2020}. 

As far as we know, the effect of stochastic resetting on a particle making Brownian motion under the logarithmic-harmonic potential defined as \linebreak $V\left(x\right)=\alpha In x + (1/2)\beta {x^2}$ has never been studied before. On the other hand, the time dependent form of the probability density function of the position of a particle diffusing with the logarithmic-harmonic potential without stochastic resetting is obtained analytically \cite{Lo2003,Giampaoli1999,Ryabov2013}. In recent years, the Logarithmic-harmonic potential has been the subject of some research in the field of stochastic thermodynamics. For example, heat fluctuations in a diffusion system are handled under the logarithmic-harmonic potential and full heat distribution is obtained \cite{Paraguassu2022}. The work distribution of a Brown particle diffusing in a logarithmic-harmonic potential is investigated and the importance of this distribution in terms of obtaining equilibrium free energy differences in experiments based on Jarzynski identity is discussed \cite{Holubec2015}. Another issue where logarithmic-harmonic potential is important in stochastic thermodynamics is stochastic heat engines. Efficiency is optimized by considering a fully solvable example of a heat engine based on such a system \cite{Holubec2014}. The logarithmic-harmonic potential has also been used in finance to model the asymmetric behavior of unemployment rates, which increase rapidly during economic recessions but gradually decrease during growth periods \cite{Hui2022}. Because of all these applications we have mentioned, our focus in this study is on a particle diffusing in a logarithmic-harmonic potential with stochastic resetting.

The present paper is organized as follows: In Section \ref{sec:section2}, We begin by considering a particle diffusing under a logarithmic-harmonic potential. We introduce the Fokker-Planck equations that give the probability distribution of the position of the diffusing particle without stochastic reset in Section \ref{subsec:subsection1} and with stochastic reset in Section \ref{subsec:subsection2}. In Section \ref{sec:section3}, we analytically obtain the steady state distribution for diffusion in the logarithmic-harmonic potential with stochastic reset. The main contribution of this article to the literature is analytically obtaining Eq. \ref{eq:9}, which gives the steady state distribution for diffusion in logarithmic-harmonic potential, and revealing the important features of the distribution, which we discussed in Figure \ref{figure3} and Figure \ref{figure4}. We conclude by summarizing the results in Section \ref{sec:section4}.

\begin{figure*}[!htbp]
\centering
    \includegraphics[width=0.47\linewidth]{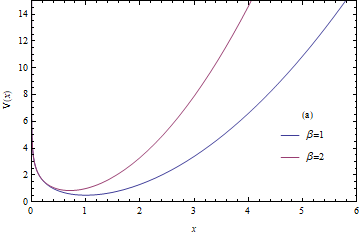}\hfil
    \includegraphics[width=0.47\linewidth]{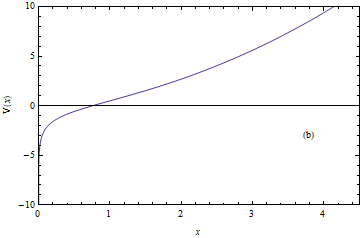}\par\medskip
    \caption{Logarithmic-harmonic potential as a function of $x$ with respect to different values of parameters. (a) $\alpha  = 1$ and $\beta  = 1$ are represented by the blue line, $\beta  = 2$ is represented by the pink line. As the value of   increases, the minimum of the potential approaches the origin. (b) $\alpha  = -1$ and $\beta  = 1$. For negative $\alpha$ values, the behavior of the potential around the origin changes.}
    \label{figure1}
\end{figure*}

\begin{figure*}[!htbp] 
\centering
    \includegraphics[width=0.47\linewidth]{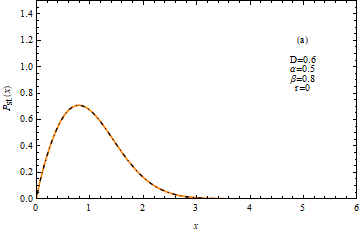}\hfil
    \includegraphics[width=0.47\linewidth]{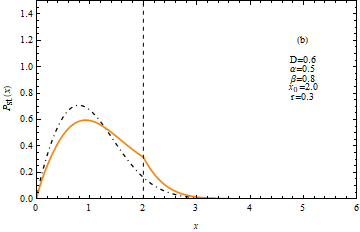}\par\medskip
    \caption{The stationary distribution function ${P_{st}}\left( x \right)$ for logarithmic-harmonic potential. The dashed black line is obtained by using Eq. \ref{eq:6}, which gives stable distribution in the absence of stochastic reset, and the orange colored line is obtained by using Eq. \ref{eq:9}, which gives stable distribution under stochastic reset. Parameters $D = 0.6, \alpha  = 0.5, \beta  = 0.8$ and (a) reset rate $r = 0$, (b) reset rate $r = 0.3$ and initial position ${x_0} = 2$}
    \label{figure2}
\end{figure*}

\begin{figure*}[!htbp]
\centering
    \includegraphics[width=0.47\linewidth]{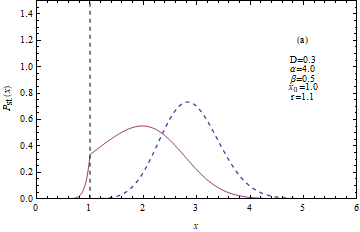}\hfil
    \includegraphics[width=0.47\linewidth]{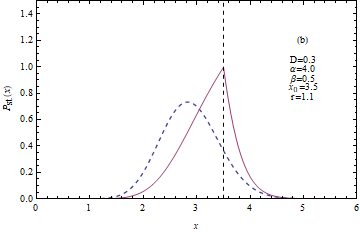}\par\medskip
    \includegraphics[width=0.47\linewidth]{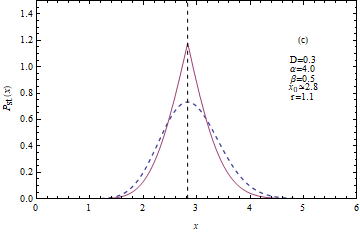}\hfil
    \includegraphics[width=0.47\linewidth]{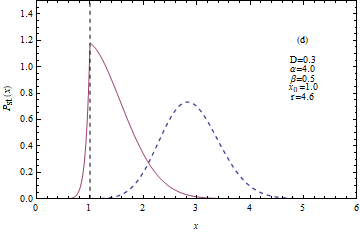}
    \caption{Stationary distribution function ${P_{st}}\left( x \right)$ for logarithmic-harmonic potential. The dashed blue line represents stable distribution without stochastic reset and sold line under stochastic reset. We chose the parameters as $D = 0.3, \alpha  = 4.0, \beta  = 0.5$. Reset rate is $r$ and reset position is ${x_0}$ (a) $r=1.1$ and ${x_0}=1.0$, (b) $r=1.1$ and ${x_0}=3.5$, (c) $r=1.1$ and ${x_0} \approx 2.8$ (minimum of potential), (d) $r=4.6$ and ${x_0}=1.0$.}
    \label{figure3}
\end{figure*}

\begin{figure*}[!htbp] 
\centering
    \includegraphics[width=0.47\linewidth]{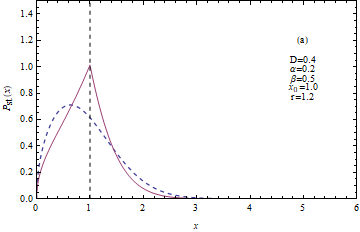}\hfil
    \includegraphics[width=0.47\linewidth]{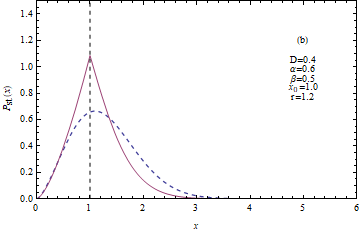}\par\medskip
    \includegraphics[width=0.47\linewidth]{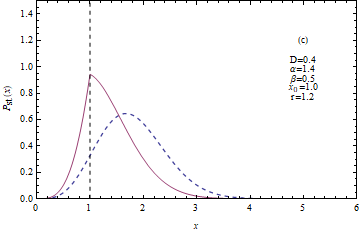}\hfil
    \includegraphics[width=0.47\linewidth]{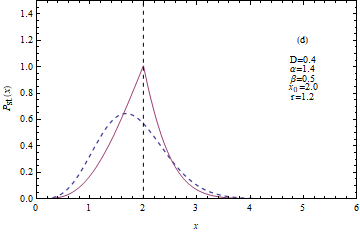}
    \caption{Stationary distribution function ${P_{st}}\left( x \right)$ for logarithmic-harmonic potential. The dashed blue line represents stable distribution without stochastic reset and solid line under stochastic reset. We chose the parameters as $D = 0.4, r=1.2, \beta  = 0.5$. We chose the parameter $\alpha$ of the logarithmic part in the potential and the reset position ${x_0}$ (a) $\alpha  = 0.2$ and ${x_0}=1.0$, (b) $\alpha  = 0.6$ and ${x_0}=1.0$, (c) $\alpha  = 1.4$ and ${x_0}=1.0$, (d) $\alpha  = 1.4$ and ${x_0}=2.0$.}
    \label{figure4}
\end{figure*}  
\section{MODEL} \label{sec:section2}
Consider the logarithmic-harmonic potential of the form 
\begin{eqnarray} \label{eq:1}
V\left(x \right) = -\alpha In x + \frac{1}{2}\beta {x^2}
\end{eqnarray}
where $x>0$ \cite{Lo2003,Giampaoli1999,Ryabov2013}. The position $x(t)$ evolution in one dimension of a single particle diffusing under an external logarithmic-harmonic potential is obtained by a Langevin equation:
\begin{eqnarray} \label{eq:2}
\frac{{dx}}{{dt}} =  - V'(x) + \eta (t) = \frac{\alpha }{x} - \beta x + \eta (t)
\end{eqnarray}
where $\eta (t)$ is a Gaussian White noise with statistical properties
\begin{eqnarray} \label{eq:3}
\left\langle {\eta (t)} \right\rangle  = 0, \left\langle {\eta (t)\eta (t')} \right\rangle  = 2D \delta (t - t')
\end{eqnarray}
and $D$ is the diffusion constant. The initial condition is $x(0) = {x_0}$ where ${x_0} \in \left( {0,\infty } \right)$. The potential has a singularity at the origin due to its logarithmic part. Certainly, in the deterministic case without stochastic effects, the particle always moves on the positive side of the $x$-axis and cannot reach the origin. On the other hand, staying of the particle on the positive side of its axis under stochastic effects depends on the value of the noise and the strength of the logarithmic part of the potential. In this paper we are only concerned with case $\alpha>0$ and $\beta>0$. As seen in Figure \ref{figure1}, choosing $\alpha$ and $\beta$ parameters positive due to the minus at the beginning of the logarithmic part of the potential ensures that the particle is always in the positive $x$ region. Parameter $\alpha$ determines the power of the logarithmic part and parameter $\beta$ determines the shape of the harmonic part that changes more slowly and creates a potential asymmetric trap.    
\begin{subsection}{Without stochastic resetting} \label{subsec:subsection1}
     If the particle is at position ${x_0}$ at time $t=0$, the Fokker-Planck equation,
\begin{eqnarray} \label{eq:4}
\frac{{\partial p\left( {\left. {x,t} \right|{x_0}} \right)}}{{\partial t}} &=& D\frac{{{\partial ^2}p\left( {\left. {x,t} \right|{x_0}} \right)}}{{\partial {x^2}}} \notag
\\ &&+ \frac{\partial }{{\partial x}}\left[ {\left( { - \frac{\alpha }{x} + \beta x} \right)p\left( {\left. {x,t} \right|{x_0}} \right)} \right]
\end{eqnarray}
obtained through the standard approximation from the Langevin equation in Eq. \ref{eq:2}, provides the evolution of the conditional probability distribution function $p\left( {\left. {x,t} \right|{x_0}} \right)$ that describes the particle's presence at position $x$ at time $t$. The initial condition is $p\left( {\left. {x,0} \right|{x_0}} \right) = \delta \left( {x - {x_0}} \right)$. Time-dependent distribution $p\left( {\left. {x,t} \right|{x_0}} \right)$ can be found using path integral approach \cite{Giampaoli1999} or Lie algebraic method \cite{Ryabov2013}. $\partial p\left( {\left. {x,t} \right|{x_0}} \right)/\partial t = 0$ is taken to get steady state solition:
\begin{eqnarray} \label{eq:5}
&D&\frac{{{\partial ^2}{p_{st}}\left( {\left. x \right|{x_0}} \right)}}{{\partial {x^2}}} + \notag \\
&&\frac{\partial }{{\partial x}}\left[ {\left( { - \frac{\alpha }{x} + \beta x} \right){p_{st}}\left( {\left. x \right|{x_0}} \right)} \right] = 0.
\end{eqnarray}
The steady state solution, where $\alpha /D \ge  - 1$ is in the form of 
\begin{eqnarray} \label{eq:6}
{p_{st}}\left( {\left. x \right|{x_0}} \right) = \frac{{{2^{\frac{{ - \alpha  + D}}{{2D}}}}{{\left( {\frac{\beta }{D}} \right)}^{\frac{{\alpha  + D}}{{2D}}}}{x^{\alpha /D}}}}{{\Gamma \left( {\frac{{\alpha  + D}}{{2D}}} \right)}}\exp \left( { - \frac{{\beta {x^2}}}{{2D}}} \right).
\end{eqnarray}
Since $\alpha>0$ is in our model, the $\alpha /D \ge  - 1$ condition is provided.
  \end{subsection}
  \begin{subsection}{With stochastic resetting} \label{subsec:subsection2}
We now consider a stochastic reset mechanism in which the particle is sent to its initial position with a constant rate of $r$. The $x$-position of the particle is reset to position $x={x_0}$ with probability $r\Delta t$ in a small time interval $\Delta t$, or with probability $1-r\Delta t$ the particle exhibits the dynamic behavior given in Eq. \ref{eq:2}. Under such a stochastic reset mechanism, the Fokker Planck equation giving the evolution of $P\left( {\left. {x,t} \right|{x_0}} \right)$ becomes:
\begin{eqnarray} \label{eq:7}
\frac{{\partial P\left( {\left. {x,t} \right|{x_0}} \right)}}{{\partial t}} &=& D\frac{{{\partial ^2}P\left( {\left. {x,t} \right|{x_0}} \right)}}{{\partial {x^2}}} \notag
\\&&+ \frac{\partial }{{\partial x}}\left[ {\left( { - \frac{\alpha }{x} + \beta x} \right)P\left( {\left. {x,t} \right|{x_0}} \right)} \right] \notag
\\&&- r P\left( {\left. {x,t} \right|{x_0}} \right) + r \delta \left( {x - {x_0}} \right).
\end{eqnarray}
The initial condition is $P\left( {\left.{x,0}\right|{x_0}}\right)=\delta\left( {x - {x_0}}\right)$. The last two terms on the right appear due to the reset mechanism. The negative term represents the reduce in probability of being in $x$ due to resetting to $x={x_0}$, while the final term represents the gain of probability of being at $x={x_0}$ due to resetting from all other positions. Thus, the steady state solution under stochastic reset is obtained by the equation $P\left( {\left. {x,t} \right|{x_0}} \right)$, 
\begin{eqnarray} \label{eq:8}
&D&\frac{{{\partial ^2}{P_{st}}\left( {\left. x \right|{x_0}} \right)}}{{\partial {x^2}}} + \frac{\partial }{{\partial x}}\left[ {\left( { - \frac{\alpha }{x} + \beta x} \right){P_{st}}\left( {\left. x \right|{x_0}} \right)} \right] \notag
\\&&- r {P_{st}}\left( {\left. x \right|{x_0}} \right) + r \delta \left( {x - {x_0}} \right) = 0.
\end{eqnarray}
\end{subsection}
\section{STEADY STATE DISTRIBUTION \\ UNDER STOCHASTIC RESET} \label{sec:section3}
Since we have $\alpha>0$ and $\beta>0$ in our model, the steady state distribution in the logarithmic-harmonic potential in the absence of stochastic resetting is centered around the $x = \sqrt {\alpha /\beta }$ minimum of the potential. On the other hand, under stochastic reset, the ${P_{st}}\left( {\left. x \right|{x_0}} \right)$ stationary distribution function (SDF) is continuous everywhere and tends to have a sharp peak in $x={x_0}$. Therefore, the derivative of the SDF at the $x={x_0}$ point is discontinuous. Thus, SDF becomes a piecewise function represented in two different regions, $x < {x_0}$ and $x > {x_0}$. General solution of Eq. \ref{eq:8} is given by
\begin{widetext}
\begin{eqnarray} \label{eq:9}
P_{st}^{ x > {x_0}}\left( {\left. x \right|{x_0}} \right) = {a_1} {x^{{\alpha  \mathord{\left/
 {\vphantom {\alpha  D}} \right.
 \kern-\nulldelimiterspace} D}}}{}_1{F_1}\left( {\frac{{\beta (\alpha  + D) - rD}}{{2\beta D}};\frac{{\alpha  + D}}{{2D}}; - \frac{\beta }{{2D}}{x^2}} \right) + {a_2}{\kern 1pt} x{\kern 1pt} {\kern 1pt} {}_1{F_1}\left( {\frac{{2\beta  - r}}{{2\beta }};\frac{{3D - \alpha }}{{2D}}; - \frac{\beta }{{2D}}{x^2}} \right), \notag
 \\ P_{st}^{ x < {x_0}}\left( {\left. x \right|{x_0}} \right) = {a_3}{\kern 1pt} {x^{{\alpha  \mathord{\left/
 {\vphantom {\alpha  D}} \right.
 \kern-\nulldelimiterspace} D}}}{}_1{F_1}\left( {\frac{{\beta (\alpha  + D) - rD}}{{2\beta D}};\frac{{\alpha  + D}}{{2D}}; - \frac{\beta }{{2D}}{x^2}} \right) + {a_4}{\kern 1pt} x{\kern 1pt} {\kern 1pt} {}_1{F_1}\left( {\frac{{2\beta  - r}}{{2\beta }};\frac{{3D - \alpha }}{{2D}}; - \frac{\beta }{{2D}}{x^2}} \right),
\end{eqnarray}
\end{widetext}
where $_1{F_1}\left( {a;b;x} \right)$ is the Kummer confluent hypergeometric function. 

The confluent hypergeometric function has a hypergeometric series given by
\begin{eqnarray} \label{eq:10}
{}_1{F_1}\left( {a;b;x} \right) = \sum\limits_{n = 0}^\infty  {\frac{{{{\left( a \right)}_n}{x^n}}}{{{{\left( b \right)}_n}n!}}} 
\end{eqnarray}
where ${\left( a \right)_n}$ and ${\left( a \right)_n}$ are Pochhammer symbols. The Kummer confluent hypergeometric function is an analytic, single-valued function of all real values of $a,b$ and $x$, except when $b$ is a nonzero or negative integer. $P_{st}^{{\kern 1pt} x < {x_0}}\left( 0 \right) = 0$ is self-contained and SDF is converged in format $\mathop {\lim }\limits_{x \to \infty } P_{st}^{{\kern 1pt} x > {x_0}}\left( x \right) = 0$.
According to the asymptotic form of the Kummer confluent hypergeometric function, \linebreak $\mathop {\lim }\limits_{x \to \infty } {}_1{F_1}\left( {a;b;x} \right) = \frac{{\Gamma \left( b \right)}}{{\Gamma \left( a \right)}}{\kern 1pt} {\kern 1pt} {e^x} {x^{a - b}}$ becomes. The term containing ${a_1}$ converges as ${x^{\left( {\frac{\alpha }{D} - \frac{r}{{2\beta }}} \right)}}\exp \left( { - \frac{\beta }{{2D}}{x^2}} \right)$ and the term containing ${a_2}$ converges as ${x^{\left( {\frac{1}{2} - \frac{r}{{2\beta }} + \frac{\alpha }{D}} \right)}}\exp \left( { - \frac{\beta }{{2D}}{x^2}} \right)$ when $x \to \infty $. Thus, $r{\kern 1pt} D > 2\alpha \beta $ when $\alpha  > D$ and $r{\kern 1pt} D > \alpha {\kern 1pt} \beta  + \beta {\kern 1pt} D$ when $\alpha  < D$ generally ensures convergence. On the other hand, since we have $\beta  > 0$ in our model, these conditions do not need to be met to ensure convergence because of the $\exp \left( { - \frac{\beta }{{2D}}{x^2}} \right)$ factor in asymptotic form. At this stage, we choose ${a_4} = 0$. Thus, after obtaining ${a_2}$ and ${a_2}$ by utilizing Eq. \ref{eq:11}, we have the opportunity to find ${a_1}$ by using the normalization condition.

We obtain, by integrating Eq. \ref{eq:8} over an infinitesimal region around $x = {x_0}$, 
\begin{equation} \label{eq:11}
{\left. {\frac{d}{{dx}}P_{st}^{{\kern 1pt} x > {x_0}}\left( {\left. x \right|{x_0}} \right)} \right|_{x = {x_0}}} - {\left. {\frac{d}{{dx}}P_{st}^{ x < {x_0}}\left( {\left. x \right|{x_0}} \right)} \right|_{x = {x_0}}} =  - \frac{r}{D}  
\end{equation} 
Thus we get,
\begin{widetext}
\begin{eqnarray} \label{eq:12}
\begin{array}{cc}
\left( {{a_1} - {a_3}} \right) \left( {\frac{\alpha }{D} {x^{\frac{\alpha }{D} - 1}}{}_1{F_1}\left( {\frac{{\beta \left( {\alpha  + D} \right) - rD}}{{2\beta D}};\frac{{\alpha  + D}}{{2D}}; - \frac{{\beta  x_0^2}}{{2D}}} \right) - \frac{{\alpha \beta  + D\left( {\beta  - r} \right)}}{{D\left( {\alpha  + D} \right)}} {x^{\frac{\alpha }{D} + 1}}{}_1{F_1}\left( {\frac{{\beta \left( {\alpha  + 3D} \right) - rD}}{{2\beta D}};\frac{{\alpha  + 3D}}{{2D}}; - \frac{{\beta  x_0^2}}{{2D}}} \right)} \right)\  \\
 \\+  {a_2}\left( {{}_1{F_1}\left( {\frac{{2\beta  - r}}{{2\beta }};\frac{{3D - \alpha }}{{2D}}; - \frac{{\beta {\kern 1pt} x_0^2}}{{2D}}} \right) + \frac{{2\beta  - r}}{{\alpha  - 3D}} x_0^2 {}_1{F_1}\left( {\frac{{4\beta  - r}}{{2\beta }};\frac{{5D - \alpha }}{{2D}}; - \frac{{\beta x_0^2}}{{2D}}} \right)} \right) =  - \frac{r}{D}.
\end{array}
\end{eqnarray}
Since SDF is continuous at $x = {x_0}$, is obtained and we find
\begin{eqnarray} \label{eq:13-14}
{a_2} = \frac{{{m_2}}}{n},
\\{a_3} = {a_1} + \frac{{{m_1}}}{n}
\end{eqnarray}
from Eq.(12) taking ${m_1} ,{m_2}$ and $n$ values below: 
\begin{eqnarray} \label{eq:15-16}
{m_1} \equiv \left( {\alpha  - 3D} \right)\left( {\alpha  + D} \right)r {x_0}^{1 - \frac{\alpha }{D}} {}_1{F_1}\left( {\frac{{2\beta  - r}}{{2\beta }};\frac{{3D - \alpha }}{{2D}}; - \frac{{\beta  x_0^2}}{{2D}}} \right),\\ \notag
\\{m_2} \equiv \left( {\alpha  - 3D} \right)\left( {\alpha  + D} \right)r {}_1{F_1}\left( {\frac{{\beta \left( {\alpha  + D} \right) - rD}}{{2\beta D}};\frac{{\alpha  + D}}{{2D}}; - \frac{{\beta x_0^2}}{{2D}}} \right),
\end{eqnarray}
\begin{eqnarray} \label{eq:17}
n &\equiv& \left( {3D - \alpha } \right)\left( {\alpha \beta  + D\left( {\beta  - r} \right)} \right) x_0^2{}_1{F_1}\left( {\frac{{\beta \left( {\alpha  + 3D} \right) - rD}}{{2\beta D}};\frac{{\alpha  + 3D}}{{2D}}; - \frac{{\beta x_0^2}}{{2D}}} \right) {}_1{F_1}\left( {\frac{{2\beta  - r}}{{2\beta }};\frac{{3D - \alpha }}{{2D}}; - \frac{{\beta x_0^2}}{{2D}}} \right)  \notag \\ 
&+& \left( {\alpha  + D} \right) {}_1{F_1}\left( {\frac{{\beta \left( {\alpha  + D} \right) - rD}}{{2\beta D}};\frac{{\alpha  + D}}{{2D}}; - \frac{{\beta  x_0^2}}{{2D}}} \right) \left[ {\left( {\alpha  - 3D} \right)\left( {\alpha  - D} \right) {}_1{F_1}\left( {\frac{{2\beta  - r}}{{2\beta }};\frac{{3D - \alpha }}{{2D}}; - \frac{{\beta  x_0^2}}{{2D}}} \right)} \right. \notag \\ 
&&\left. { +  D \left( {r - 2\beta } \right)x_0^2{}_1{F_1}\left( {\frac{{4\beta  - r}}{{2\beta }};\frac{{5D - \alpha }}{{2D}}; - \frac{{\beta x_0^2}}{{2D}}} \right)} \right].
\end{eqnarray}
 ${a_1}$ is found from 
 \begin{eqnarray} \label{eq:18}
 \int\limits_0^{{x_0}} {P_{st}^{x < {x_0}}\left( {\left. x \right|{x_0}} \right)} dx +  \int\limits_{{x_0}}^\infty  {P_{st}^{ x > {x_0}}\left( {\left. x \right|{x_0}} \right)} dx = 1
 \end{eqnarray}
 normalization condition. Thus, the steady state distribution of the position of a particle exposed to the stochastic reset mechanism in logarithmic-harmonic potential is obtained by Eq. \ref{eq:9}.
\end{widetext}
Using Eq. \ref{eq:6} in the absence of stochastic resetting, we plotted the SDF shaped by the structure of the potential for $D = 0.6,\alpha  = 0.5, \beta  = 0.8$ values in Figure \ref{figure2}a (The dashed black line).

In addition, we plotted the Eq. \ref{eq:9} we found for a Brownian particle exposed to the stochastic reset mechanism by choosing $r=0$ with other parameters remaining the same (orange line). It is seen in Figure \ref{figure2}a that Eq. \ref{eq:9}, which we found for the steady state distribution of a Brownian particle exposed to stochastic reset mechanism, is reduced to Eq. \ref{eq:6}, which is valid without stochastic reset mechanism, when the reset rate is $r=0$. In Figure \ref{figure2}b, we have drawn the SDF for $D = 0.6,\alpha  = 0.5, \beta  = 0.8$ values again by using Eq. \ref{eq:6} (The dashed black line). We also plotted the SDF using Eq. \ref{eq:9} for a small zero rate of $r=0.3$ and reset position ${x_0} = 2$ (orange line). Figure \ref{figure2}b shows how the SDF is affected by sending the particle to a point ${x_0} = 2$ away from the minimum value of the potential with a small reset probability.

In Figure \ref{figure3}, the dashed blue line represents the stable distribution that occurs without stochastic reset mechanism, and the solid line represents the stable distribution that occurs under the stochastic reset mechanism. We plot the SDF using Eq. \ref{eq:6} without stochastic reset and Eq. \ref{eq:9} under stochastic reset. We chose $D = 0.3,\alpha  = 4.0, \beta  = 0.5$ in all graphs. Although the resetting ratios in Figures \ref{figure3}a and \ref{figure3}b are the same $(r=1.1)$, the characteristics of the distribution vary considerably depending on whether is to the right $({x_0} = 3.5)$ or left $({x_0} = 1.0)$ of the potential minimum. It is seen in Figure \ref{figure3}a that the SDF has a sharp structure in $x$ and does not reach its maximum value. Stochastic reset causes the maximum of the distribution centered around the minimum of the potential to shift towards ${x_0}$ and decrease its maximum. On the other hand, SDF peaks at ${x_0}$ when it is chosen to stay on the right side of the potential as seen in Figure \ref{figure3}b. Moreover, for the same reset ratio, the maximum value of the distribution increases and shifts to the right. In Figure \ref{figure3}c, we choose ${x_0}$ around the minimum of the potential. The stochastic reset mechanism causes the distribution to sharpen and peak around the potential minimum, as expected. In Figure \ref{figure3}c, the maximum value of the SDF becomes greater than in \ref{figure3}a and \ref{figure3}b. The reset position in Figure \ref{figure3}d is the same as in Figure \ref{figure3}a and the effect of the increase in $r$ on the shape of the distribution is seen. An increase in the reset rate causes the distribution to maximize at ${x_0}$.

Figure \ref{figure4} shows the effect of stochastic resetting on the distribution according to different values of $\alpha$. The ${m_1}$ and ${m_2}$ expressions in the SDF contain the $\left( {\alpha  - 3D} \right)\left( {\alpha  + D} \right)$ multipliers and $n$ includes the $\left( {\alpha  - D} \right)$ multiplier in one term. Therefore, $\alpha  =  - D, \alpha  = D,\alpha  = 3D$ are critical values in which the ${a_i}$ expressions $P_{st}^{{\kern 1pt} x > {x_0}}\left( {\left. x \right|{x_0}} \right)$ or $P_{st}^{{\kern 1pt} x < {x_0}}\left( {\left. x \right|{x_0}} \right)$ contain and some terms in these expressions change sign. We mentioned that the $\alpha /D \ge  - 1$ condition guarantees the existence of a steady state solution, while there is no stochastic reset mechanism in Section II A. Under stochastic reset in logarithmic-harmonic potential we find two new conditions that affect the structure of the SDF: $\alpha /D < 1$ and $1 < \alpha /D < 3$. We chose $D = 0.4,r = 1.2,\beta  = 0.5$ in Figure \ref{figure4}. We choose $\alpha /D = 0.5$ in Figure \ref{figure4}a, $\alpha /D = 1.5$ in \ref{figure4}b, and $\alpha /D = 3.5$ in \ref{figure4}c.The position of ${x_0}$ relative to the minimum of the potential and the evolution of the SDF in both regions are noteworthy. As seen in Figure \ref{figure4}c, when $\alpha /D > 3$, the minimum of the potential stays on the right side of ${x_0}$, while the SDF takes its maximum value at a point in the $x > {x_0}$ region, not at ${x_0}$. This is more evident in Figure \ref{figure3}a, since it has a large value such as $\alpha /D \approx 13.3$. We observe that the behavior of the SDF around ${x_0}$ changes depending on the $\alpha /D$ ratio. On the other hand, when we choose the position of ${x_0}$ on the right side of the potential minimum, the trend we mentioned is preserved according to the $\alpha /D$ ratio, but the shape of the distribution changes as seen in Figure \ref{figure4}d.
\section{SUMMARY} \label{sec:section4}
In this study, we analytically obtained the steady state distribution of the position of a Brownian particle diffusing at logarithmic-harmonic potential with the stochastic reset mechanism. We discussed the effect of resetting on the SDF by comparing the steady state distribution due to the nature of the potential in the absence of stochastic reset and the steady state distribution occurring under stochastic reset. In Figure \ref{figure3}, it is seen that the SDF exhibits different properties depending on whether the stochastic reset position ${x_0}$ is to the right or left of the potential minimum. The size of the reset rate $r$ is also an important factor determining the structure of the SDF. It was also shown that there are two critical situations, $\alpha /D < 1$ and $1 < \alpha /D < 3$, in which significant differences in the behavior of the SDF occur. The change in the structure of the SDF in response to each situation is seen in Figure \ref{figure4}. 

In our model, we chose the control parameters of the potential as positive. On the other hand, when the control parameters of the potential are not constrained in this way, the stochastic reset can be expected to have complex effects on the evolution of the system. First of all, the sign of the control parameters of the potential determines whether the origin is repulsive or attractive. Therefore, whether some new critical $\alpha /D$ values will emerge in the phase diagram, especially in terms of the first passage properties of the system, may be a triggering question for new studies.
\\
\section*{COMPETING INTERESTS} \label{sec:competing}
    The author declares no competing interests.

\bibliography{main}

\end{document}